\documentclass{easychair}
\usepackage[utf8]{inputenc}
\usepackage{xcolor}
\usepackage{xspace}
\usepackage{url}
\usepackage{xparse}
\usepackage{pbox,calc} 
\usepackage{multirow}
\usepackage{subcaption} 
\usepackage{listings}
\usepackage[font=scriptsize, labelfont=bf]{caption}
\title{\ourtitle}
\date{}
\author{
Giorgio Piras\inst{1}
\and
Maura Pintor\inst{1, 2}
\and
Luca Demetrio\inst{1, 2}
\and
Battista Biggio\inst{1, 2}
}

\institute{
  University of Cagliari,
  Cagliari, Italy\\
  \email{}
\and
   Pluribus One S.r.l.,
   Cagliari, Italy\\
   \email{}
 }

\authorrunning{Piras et al.}

\titlerunning{\ourtitle}

\begin{document}
\newcommand{\diff}[2]{\frac{\partial #1}{\partial #2}}
\newcommand{\vct}[1]{\ensuremath{\boldsymbol{#1}}}
\newcommand{\mat}[1]{\ensuremath{\mathbf{#1}}}
\newcommand{\set}[1]{\ensuremath{\mathcal{#1}}}
\newcommand{\con}[1]{\ensuremath{\mathsf{#1}}}
\newcommand{\tsum}{\ensuremath{\textstyle \sum}}
\newcommand{\T}{\ensuremath{\top}}

\definecolor{codegreen}{rgb}{0,0.6,0}
\definecolor{codegray}{rgb}{0.5,0.5,0.5}
\definecolor{codepurple}{rgb}{0.58,0,0.82}
\definecolor{backcolour}{rgb}{0.95,0.95,0.92}

\lstset{frame=tb,
  language=Python,
  aboveskip=3mm,
  belowskip=3mm,
  showstringspaces=false,
  columns=flexible,
  basicstyle={\small\ttfamily},
  numbers=none,
  backgroundcolor=\color{backcolour},   
  commentstyle=\color{codegreen},
  keywordstyle=\color{magenta},
  numberstyle=\tiny\color{codegray},
  stringstyle=\color{codepurple},
  breaklines=true,
  breakatwhitespace=true, 
  numbers=left,
  tabsize=3
}

\newcommand{\mycomment}[1]{\textcolor{olive}{Giorgio: #1}}
\newcommand{\adcomment}[1]{\textcolor{red}{Ambra: #1}}
\newcommand{\ldcomment}[1]{\textcolor{blue}{Luca: #1}}
\newcommand{\bbcomment}[1]{\textcolor{orange}{Battista: #1}}
\newcommand{\mpcomment}[1]{\textcolor{cyan}{Maura: #1}}
\newcommand{\ascomment}[1]{\textcolor{yellow}{Angelo: #1}}
\newcommand{\frcomment}[1]{\textcolor{yellow}{Fabio: #1}}

\newcommand{\ind}[1]{\ensuremath{\mathbbm 1_{#1}}}
\newcommand{\argmax}{\operatornamewithlimits{\arg\,\max}}
\newcommand{\erf}{\text{erf}}
\newcommand{\sign}{\text{sign}}
\newcommand{\argmin}{\operatornamewithlimits{\arg\,\min}}
\newcommand{\bmat}[1]{\begin{bmatrix}#1\end{bmatrix}}

\newcommand{\myparagraph}[1]{\smallskip \noindent \textbf{#1}}
\newcommand{\ie}{{i.e.}\xspace}
\newcommand{\eg}{{e.g.}\xspace}
\newcommand{\etal}{{et al.}\xspace}
\newcommand{\etc}{{etc.}\xspace}
\newcommand{\aka}{{a.k.a.}\xspace}
\newcommand{\wrt}{{w.r.t.}\xspace}

\newcommand{\ellzero}{$\ell_0$\xspace}
\newcommand{\ellone}{$\ell_1$\xspace}
\newcommand{\elltwo}{$\ell_2$\xspace}
\newcommand{\ellinf}{$\ell_{\infty}$\xspace}

\newcommand{\arecord}{\texttt{A}\xspace}
\newcommand{\nsrecord}{\texttt{NS}\xspace}
\newcommand{\ptrrecord}{\texttt{PTR}\xspace}
\newcommand{\nxdomain}{\texttt{NXDOMAIN}\xspace}
\newcommand{\pcap}{\texttt{.pcap}\xspace}

\newcommand{\uniqueips}{\texttt{unique\_ips}\xspace}
\newcommand{\lms}{\texttt{\%of\_lms}\xspace}
\newcommand{\numchars}{\texttt{num\_chars\%}\xspace}

\newcommand{\ourtitle}{Explaining Machine Learning DGA Detectors from DNS Traffic Data}

\newcommand{\maxpages}{10}
\ExplSyntaxOn
\fp_new:N \g_total_sum_fp
\NewDocumentCommand{\AddValue}{m}{
  \fp_gadd:Nn \g_total_sum_fp {#1}
}
\NewDocumentCommand{\DisplaySum}{}{
  \fp_to_decimal:N \g_total_sum_fp
}
\ExplSyntaxOff
\newcommand{\budget}[1]{\textcolor{red}{max #1 pages \AddValue{#1}[\DisplaySum/\maxpages \xspace pages]\\}}

\maketitle

\begin{abstract}
One of the most common causes of lack of continuity of online systems stems from a widely popular Cyber Attack known as Distributed Denial of Service (DDoS), in which a network of infected devices (botnet) gets exploited to flood the computational capacity of services through the commands of an attacker. This attack is made by leveraging the Domain Name System (DNS) technology through Domain Generation Algorithms (DGAs), a stealthy connection strategy that yet leaves suspicious data patterns. To detect such threats, advances in their analysis have been made. For the majority, they found Machine Learning (ML) as a solution, which can be highly effective in analyzing and classifying massive amounts of data. Although strongly performing, ML models have a certain degree of obscurity in their decision-making process. To cope with this problem, a branch of ML known as Explainable ML tries to break down the black-box nature of classifiers and make them interpretable and human-readable. This work addresses the problem of Explainable ML in the context of botnet and DGA detection, which at the best of our knowledge, is the first to concretely break down the decisions of ML classifiers when devised for botnet/DGA detection, therefore providing global and local explanations.  
\end{abstract}

\section{Introduction}
During the last decades, our day-by-day life has been strictly connected to the usage of devices and online services, therefore making their efficiency and continuity play a crucial role in the technological transformation we witness. 
Likewise, the economic loss derived from cyber-threats has increased exponentially in recent years~\cite{smith_hidden_2020} as the technologies continually evolve and attackers develop their skills. One of the most common ways cybercriminals try to jeopardize the continuity of systems and thus cause economic damage is Denial of Service (DoS), which aims to drain the computing capabilities of the target system in both fancy and basic ways. A case of this attack is the Distributed Denial of Service DDoS, where a network of infected devices (bots) are commanded by an attacker (botmaster) through a Command\&Control Server (C\&C)~\cite{puri_bots_2003, salusky_know_2007, stone-gross_your_2009}.  
What happens to be erratic and thus detectable by a Machine Learning (ML) model in this kind of attack is the DNS traffic, carrying Domain Names through which bots are connected to the C\&C server. This stealthy connection strategy is commonly known as Domain Fluxing, where the algorithms used by the infected bots to generate the domain are known as Domain Generation Algorithms (DGAs).

Although employing ML models to detect the presence of botnets within network traffic has been demonstrated to be successful, almost the entirety of the relevant works have followed a common baseline and workflow, presenting a partially novel feature set on which to train a classifier to obtain relevant results~\cite{bilge2011exposure, Schueppen2018, zheng_themis_2019, schiavoni2014phoenix}. The proposed approaches lack interpretability and contextualization. First, depending on the context from which DNS traffic data is extracted and the model is deployed, potential attackers might have control over some features. Second, the model prioritization and general usage of the features in the decision process are not known beforehand, making the process challenging to debug and protect. 

To make up for these problems, we first analyze the techniques used to detect botnets/DGAs from the DNS data (Section~\ref{sec:dns-system}); we analyze which explainability techniques can provide insight into how the model takes its decisions (Section~\ref{sec:explainability}). Upon a re-implementation of the EXPOSURE system~\cite{bilge2014exposure, bilge2011exposure} (Section~\ref{sec:experiments}), we provide the following contributions:  
(i) we build and test the EXPOSURE system on a newly collected dataset; (ii) we observe statistics on the features used by the system; (iii) we train different classifiers and compare their performances; (iv) we obtain explanations from such classifiers; and (v) given the explanations, we develop and discuss an analysis on the features used by the systems mentioned above.
Finally, we conclude the work by presenting related works (Section~\ref{sect:related_works}), limitations, and future directions (Section~\ref{sec:conclusions}).

\section{Background: DNS System and ML Techniques}\label{sec:dns-system} 
\myparagraph{From DNS to DGA.} The Domain Name System (DNS) is a database responsible for mapping domain names to IP addresses, thus answering a query made by clients in the form of a domain name towards the IP addresses. This action is commonly known as \textit{resolution}~\cite{mockapetris_domain_nodate}. 
The DNS organizes domain names into a hierarchy (through dot-separated levels), as the whole technology itself creates a hierarchical database structure. 
The information stored and carried by DNS records can be \arecord records, returning IPv4 addresses, \nsrecord records, returning authoritative name servers, and finally, \ptrrecord, which stands for Pointer to Record and returns a domain name but in the reverse query format (\ie, the question started from an IP rather than a domain name). It is also worth citing specific information carried by the DNS packets, such as the Time-to-live (TTL), which indicates how long the server will cache that packet~\cite{mockapetris_domain_nodate-1, mockapetris_domain_nodate}. Being of paramount importance for the correct functioning of basic internet activities, the DNS is the perfect target for malicious activities having a high impact on unaware users. 
That is why this technology gets exploited by attackers (\textit{botmasters}) who aim to command and control a network of infected machines, \ie, a botnet.
To go as undercover as possible, having only one domain name to which to connect would have the botnet quickly taken down by vigilant authorities. That is why bots generate massive DNS traffic trying to connect to a much more concealed C\&C server.
The generation of such a significant amount of domain names happens through Domain Generation Algorithms that, given a random seed, create a string that will possibly establish a connection. 

\myparagraph{Botnet Detection with ML: The EXPOSURE system.} Seizing the chance to detect malicious patterns, the research community has driven its efforts towards analyzing the DNS data, extracting the features, and eventually training a ML model capable to distinguish malicious and benign DNS behaviors. The EXPOSURE system~\cite{bilge2014exposure, bilge2011exposure} is among the most prominent works for its completeness in the feature set and reproducibility (in terms of feature extraction). For this reason, we use it as a base for our explainability analysis.
Table~\ref{tab:exp_features_mnem} shows the feature set, listing the features extracted by the EXPOSURE system (whose extensive description can be found in the original work~\cite{bilge2014exposure, bilge2011exposure}). 
The entire set is subdivided into Time-Based features (collecting temporal patterns from the queries to the domains), DNS-Answer-Based Features (patterns from the answers records), TTL-Value-Based Features (statistical patterns from the TTL values), and finally, Name-Based Features (statistical patterns from the Domain Name). 
Given a collection of DNS packets, we can compose a training set of benign and malicious samples to train the classifier, as Bilge et al. did in EXPOSURE. 
Our work will focus on reproducing the experiment with our newly-collected traffic and applying explainability techniques to understand the patterns employed by the model for detecting malicious activities.

\begin{table*}
\scriptsize
\begin{center}
\begin{tabular}{| p{4cm} | p{0.5cm} | p{4cm} | p{3.4cm}|} 
\hline 
Feature set & \# & Paper Feature Name & Our Feature Names\\
\hline
\multirow{4}{*}{Time-Based Features} & 2 & Short Life & glob\_short\_lived \\   
& & & glob\_life\_ratio\\\cline{2-4} 
& 1 & Daily similarity & daily\_similarity \\\cline{2-4}  
& 2 & Repeating patterns & local\_numOf\_changes\\  
& & & stddev\_before\_change\\\cline{2-4} 
& 2 & Access Ratio & idle\\ 
& & & popular\\\cline{2-4} 
\hline
\multirow{4}{*}{DNS Answer-Based Features} & 1 & Number of distinct IP addresses & unique\_ips\\\cline{2-4}  
& 1 & Number of distinct countries & unique\_ccode\\\cline{2-4} 
& 3 & Reverse DNS query result & rev\_arec\\  
& & & rev\_nsrec\\ 
& & & rev\_asnrec\\\cline{2-4} 
& 1 & Number of domains sharing the same IP & shared\_ips\\\cline{2-4}
\hline 
\multirow{5}{*}{TTL Value-Based Features} & 1 & Average TTL & ttl\_avg\\\cline{2-4}  
& 1 & Standard Deviation of TTL & ttl\_stddev\\\cline{2-4} 
& 1 & Number of distinct TTL values & unique\_ttls\\\cline{2-4} 
& 1 & Number of TTL changes & ttl\_changes\\\cline{2-4} 
& 5 & Percentage usage of TTL ranges & ttl\_range1\\ & & & ttl\_range100\\& & & ttl\_range300\\& & & ttl\_range900\\
& & & ttl\_rangeinf\\\cline{2-4} 
\hline 
\multirow{2}{*}{Domain Name-Based Features} & 1 & \% of numerical characters & num\_chars\%\\ \cline{2-4}
& 1 &  \% of length of the LMS & \%of\_lms \\ 
\hline
\end{tabular}
\caption{List of features used in EXPOSURE renewed with a mnemonic. The first column indicates the Feature subset. The second one shows the number of features that a specific feature holds. Finally, the third and fourth columns indicate respectively the feature name chosen by the authors and our feature names for atomic features.}
\label{tab:exp_features_mnem}
\end{center}
\end{table*}

\section{Explaining Predictions of ML-based DNS Analysis} \label{sec:explainability}

As pointed out by Miller et al.~\cite{miller_explanation_2018}, explanations increase transparency and interpretability so that user awareness and systems designers can jointly benefit from this gain of trust. 
In security-relevant scenarios, like the one we are considering, understanding the data and the model provides the added benefit of helping to see if there are problems in the system, for example assigning high relevance to spurious features that should not influence it to that extent~\cite{arp2022and}.

Analyzing the dataset's statistics provides further insights into the separability of the features into the two different classes. 
Additionally, in the case under investigation, lots of features come from similar sources and elaborations, which lets the statistical analysis come in handy to highlight correlations and redundancies. 

On top of that, we will use a ML model to analyze such features and categorize the samples into the two output classes of \textit{benign} and \textit{malicious} domains.
Model explanations can help understand how the model is making such decisions. 
An explanation is said to be local if it is made on single samples and wants to describe how a model emphasizes the features of a specific single sample in its classification. On the other hand, global explanations are made over entire datasets or relevant collections of samples to describe how the model prioritizes features over those samples~\cite{molnar_25_nodate}. This work will focus on both local and global explanations.

In~\cite{lundberg_unified_2017}, Lundberg and Lee proposed SHAP (SHapley Additive exPlanations), where feature importance is computed with an additive approach, representing a unified measure of feature importance. 
The basic concept behind SHAP comes from Shapley values and a game theory setting, where the features act as players and cooperate in a coalitional game (\ie, the prediction task) to receive a profit (\ie, a gain, which is the actual prediction). 
The Shapley values assign payouts to players depending on their contribution to the total payout~\cite{hart_shapley_2017}. Thus each feature that contributes to the prediction task is computed as a sum of the expected marginal contributions in any feature value combination. Given the computational burden for which SHAP should find all the possible feature combinations, Lundberg and Lee proposed a Shapley kernel that produces estimates instead of exact values. 

\subsection{Data Analysis and Explainability Techniques} 

This section will briefly explain the details and differences between the data analysis and explainability tools that will play a central role in the experiments section. 

\myparagraph{Feature Statistical Analysis.} 
These plots show the marginal distributions of every pair of features as density plots, describing how the distributions for the classes behave. 
Through the scattered plots instead, we can assess where both benign and malicious samples lie in their ad-hoc feature space, thus making us capable of understanding to which extent pairs of features separate the data. Analyzing the scattered plots allows observing the distribution of the features to get a rough idea of how they will behave/discriminate and to which extent.

\myparagraph{Partial Dependence Plot.} This plot shows the marginal effect that a single feature has on the prediction made by the model, thus providing global explanations. 
Taking as input the model, the feature, and a background distribution on which to make the model learn the feature importance, the Partial Dependence Plots (PDPs) depict the feature values on the x-axis, whilst the y-axis represents the expected prediction contribution given the feature value. In the background, a histogram shows the underlying data distribution of the feature values. A horizontal line represents the expected contribution to the prediction, and a vertical one represents the expected value of the feature. By reading this plot, we can measure how the observed feature contributes to the classification of the samples.

\myparagraph{Summary Plot.} The SHAP summary plot, which as the PDP is a global explanation technique, shows how the model prioritizes the features and how these contribute to steering the classification towards each class. This plot comprises a list of features ordered from the one giving the higher contribution to the least powerful as interpreted by the model, showing the magnitude for benign (in blue) and malicious (in red) samples.

\myparagraph{Force Plot.}
This technique is one of the local explainability methods provided in SHAP. It explains why a specific sample has been assigned a particular label. This can be useful for understanding why samples are misclassified and to which extent the classifier misunderstands them. Force plots, showing the magnitude of the feature contribution on single samples, are rendered as blue arrows indicating magnitude values towards the benign class and red vice-versa. 

In the next section, we will use the presented techniques to explain predictions of our re-implementation of the EXPOSURE system.

\section{Experiments}\label{sec:experiments}
In the experimental section of this work, we will first describe our re-implementation of the EXPOSURE system. Then we will discuss the DNS traffic data we used to make our feature extraction, followed by a brief model selection made to improve the system's performance. Eventually, we will delve into the results section to show how explanations applied in this context can bring the analysis to the next level. 

\subsection{Re-implementation of the EXPOSURE system}

\begin{figure}[htbp] 
\centering
\includegraphics[width=\linewidth]{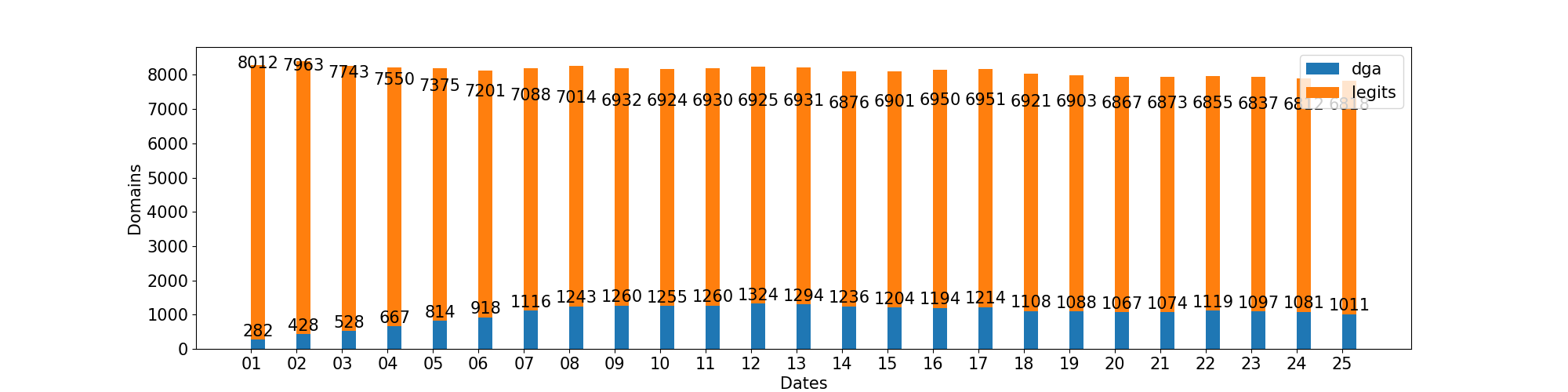}
\caption{Malicious and legitimate domains for every day.}
\label{fig:everydaydata} 
\end{figure}

\myparagraph{Dataset.} The DNS traffic was collected from recursive servers on which, through sniffers, we were able to save the data as \pcap files for the entire month of January 2021. 
Given the massive amount of traffic, summing up to 15 GB of data per day, we filtered out packets whose label was not known by either black or white lists and domain names that did not resolve (\nxdomain as response code). 
We used the list of most popular suffixes from the Alexa website\footnote{\url{https://www.alexa.com/topsites}} to label benign domains, and the list from DGArchive~\cite{plohmann2016comprehensive} to flag the malicious samples.
The remaining packets ($203,034$ domains, of which $25,882$ benign and $177,152$ malicious - note that benign domains re-appear much more frequently than the malicious ones, which are almost always unique) have then been passed through the feature extractor we implemented, and are distributed through days as shown in Figure~\ref{fig:everydaydata}.

\begin{figure}[htbp]  
\centering 
\subfloat[Classifiers trained on day 14, with class distribution unbalanced but less evident than the other days in the dataset.]{\label{fig:roc1}\includegraphics[width=0.45\linewidth]{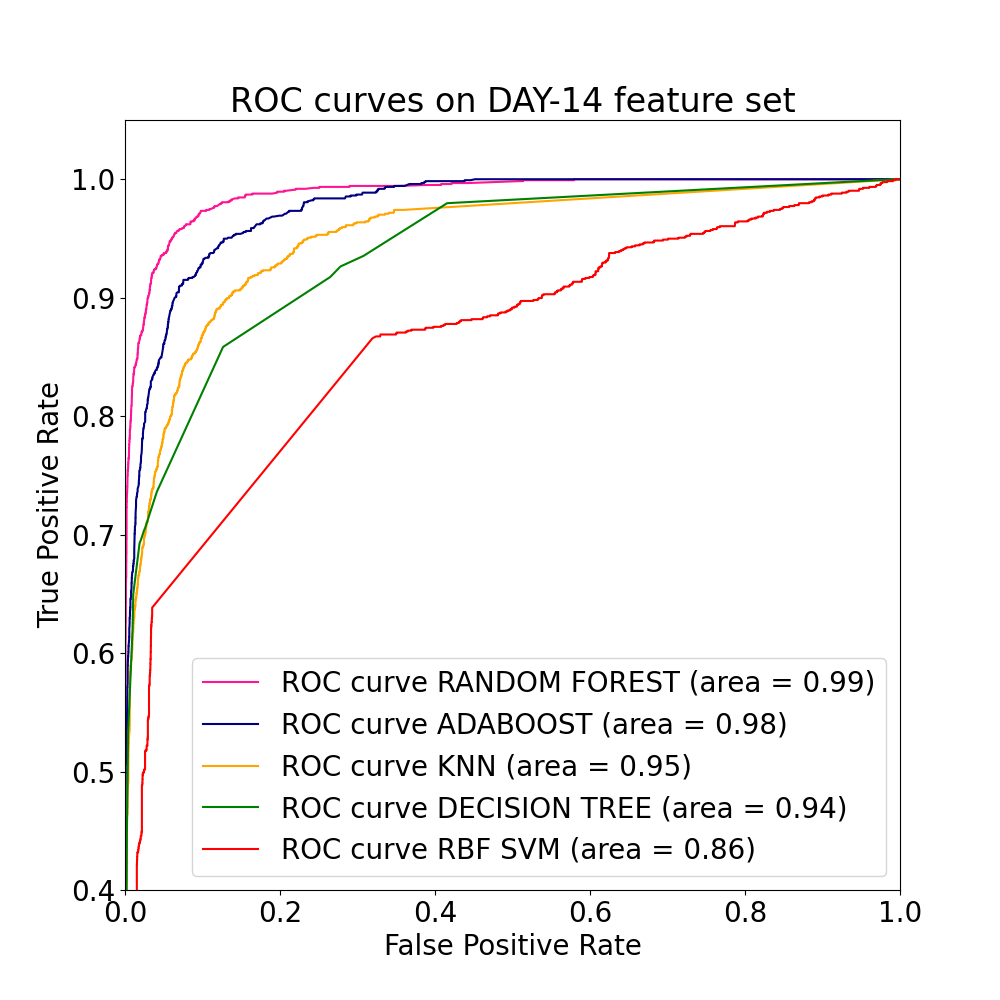}}
\qquad
\subfloat[The same classifiers trained on day 5, which presents a highly-unbalanced distribution of the two classes.]{\label{fig:roclow2}\includegraphics[width=0.45\linewidth]{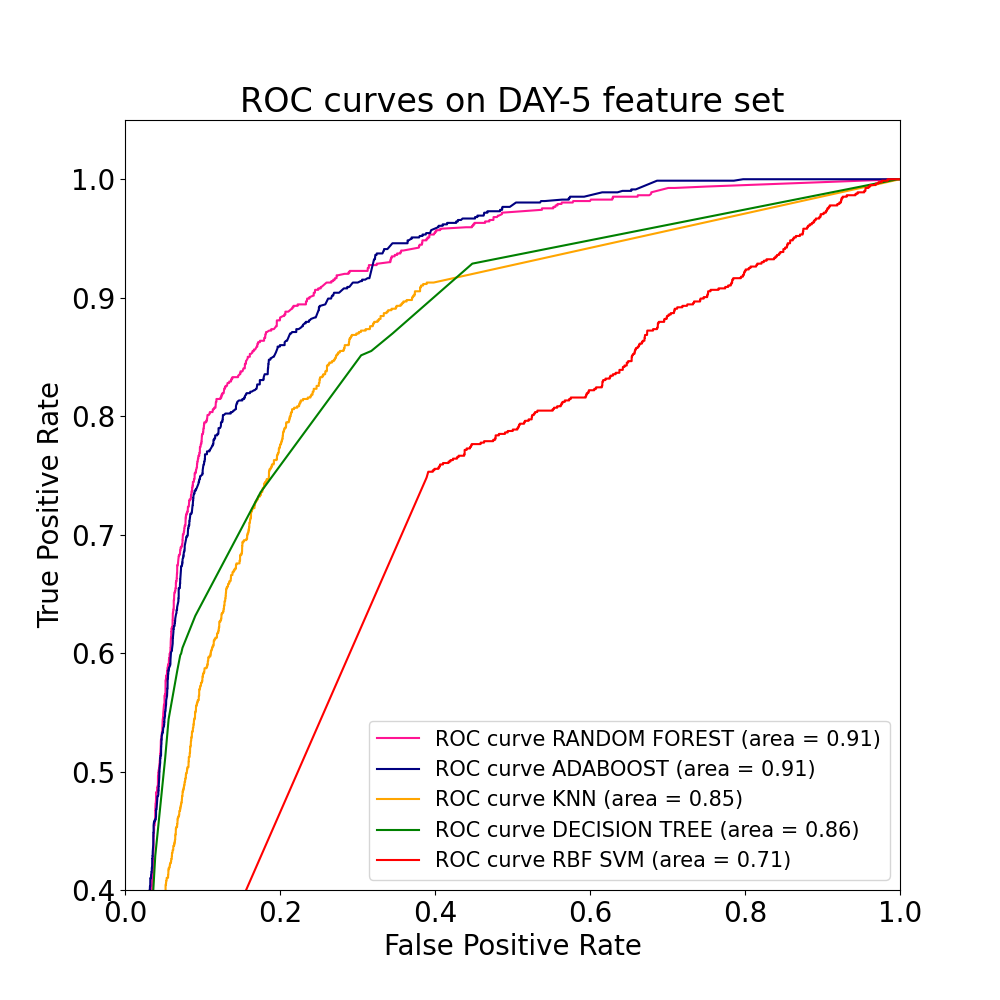}} 
\caption{The ROC curves obtained using features from day 14 (\textit{left}) and on day 5 (\textit{right}).}
\end{figure}

\myparagraph{Model Selection.}  
The authors used a J48 Decision Tree to obtain overall good performances in the original EXPOSURE work. 
We additionally bench-marked several models such as Decision Tree (DT), K-Nearest Neighbors (KNN), Support Vector Machine (SVM), Ada-Boost (ADA), and Random Forest (RF). 
After estimating the hyperparameters through a Grid Search, (whose bests overall have been reported in Section~\ref{sect:appendix}) we compared the best models with the best parameters on two different days (\ie, two different sample balances). 
The first ROC curve in Figure~\ref{fig:roc1} was obtained using a more balanced day of data (mid of January days). The second set of curves in Figure~\ref{fig:roclow2} was obtained from a day of data with very few malicious samples, showing how the performances of the classifiers dropped down consistently.
Overall, throughout the most balanced days, RF and ADA have shown to be the more consistent classifiers. For this reason, they have been selected as classifiers for the rest of the experiments. We reckon that the mid-days of capture are also more suitable for the rest of the analysis, and they have been thereby used for all of the following experiments.

\subsection{Feature and Explainability Analysis on EXPOSURE}\label{sect:results}

We now present our results on the analysis of the feature statistics and our insight obtained through applying explainability techniques to interpret the decisions taken by the machine-learning models used in our DGA detector. These proposed plots have been implemented through the Python libraries Seaborn~\cite{Waskom2021} and SHAP~\cite{noauthor_introduction_nodate}.

\begin{figure}[htbp]  
\centering 
\subfloat[Correlation between name-based features. ]{\label{fig:seaborn4}\includegraphics[width=0.45\linewidth]{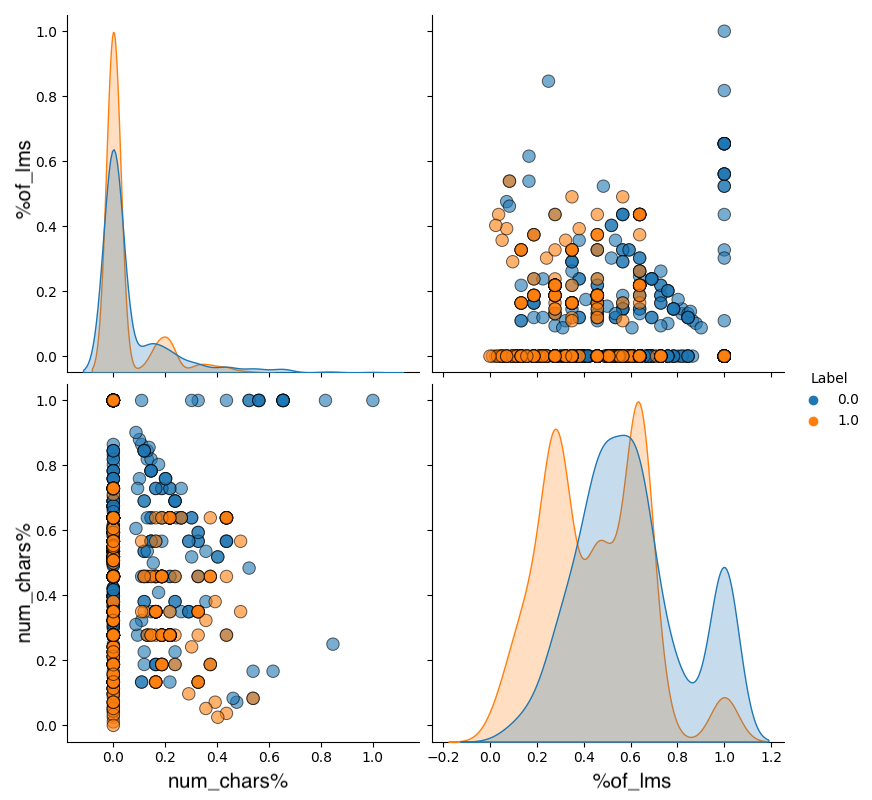} }
\qquad
\subfloat[Correlation between time-based features.]{\label{fig:seaborn_ft3}\includegraphics[width=0.45\linewidth]{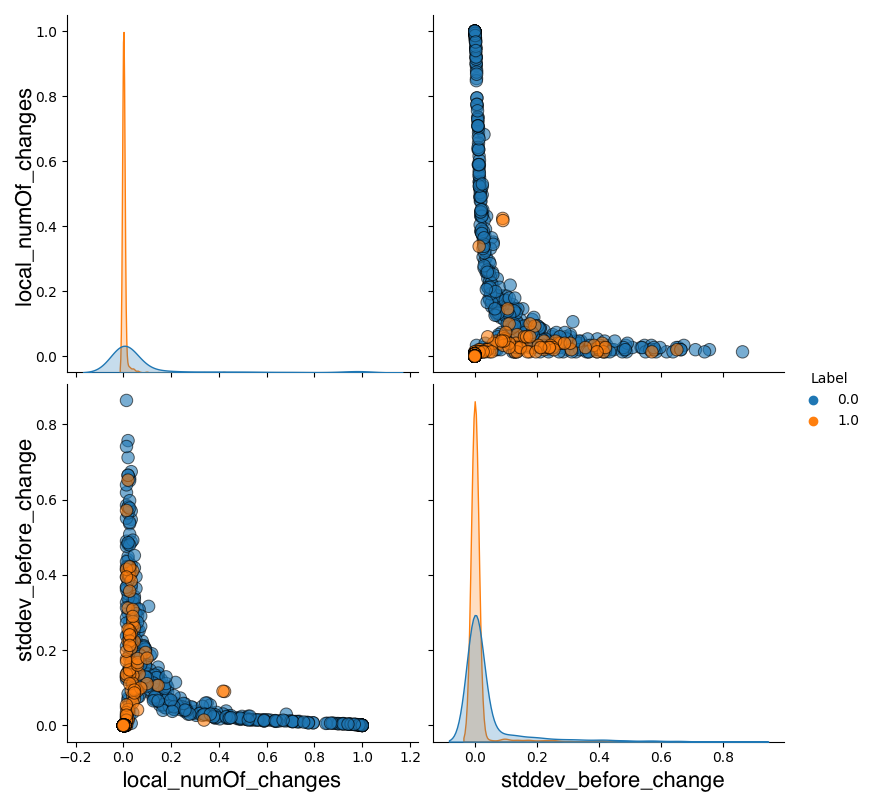}} 
\caption{Density and scatter plots for two pairs of name-based (\textit{left}) and time-based features (\textit{right}). The plots are based on the 12\textsuperscript{th} day of capture, which counts 1324 malicious samples and 1324 benign samples (resized from 6925).}
\end{figure}

\myparagraph{Statistical Analysis.} The statistical analysis shows an overview of the correlation and distribution of the features. 
 As shown in Figure~\ref{fig:seaborn4}, some features like the \lms and \numchars when joined, do not separate perfectly the data collection used into the two classes. In particular, the \lms reaches a plateau in malicious domains once over 0.8 (bottom-right plot, depicting the distribution of the feature), which describes how algorithmically-generated domains tend not to have a single meaningful word covering their entire name in most of the cases, yet there are exceptions in any direction. 
 This might be brought on by the diversity in malware families, where some like ``Gameover" DGA used to mix up numbers and characters. In contrast, others such as ``Gozi" used to mix up words from openly accessible documents, such as the US constitution~\cite{Plohmann}. 
 In the plot of Figure~\ref{fig:seaborn_ft3}, depicting time-based features, malicious domains show a more volatile behavior, which is reasonable if we think about the diversity of applications in which they can be used. 
In Figure~\ref{fig:seabornttl}, shown in the Appendix, we can observe interesting TTL behaviors characterizing the domains. Contrary to the now old-fashioned belief that a low TTL is only typical for malicious domains~\cite{4446410, alieyan_survey_2017}, as it makes malicious records stand less in caches, we show that also benign domains can present this behavior depending on the application in which they are used, \eg, to handle critical resources~\cite{vlajic_role_2012} or for load balancing purposes~\cite{alieyan_survey_2017}.  

\begin{figure}[htbp]  
\centering 
\subfloat[SHAP summary plot of feature contributions on ADA BOOST classifier.]{\label{fig:shap_ada}\includegraphics[width=0.475\linewidth]{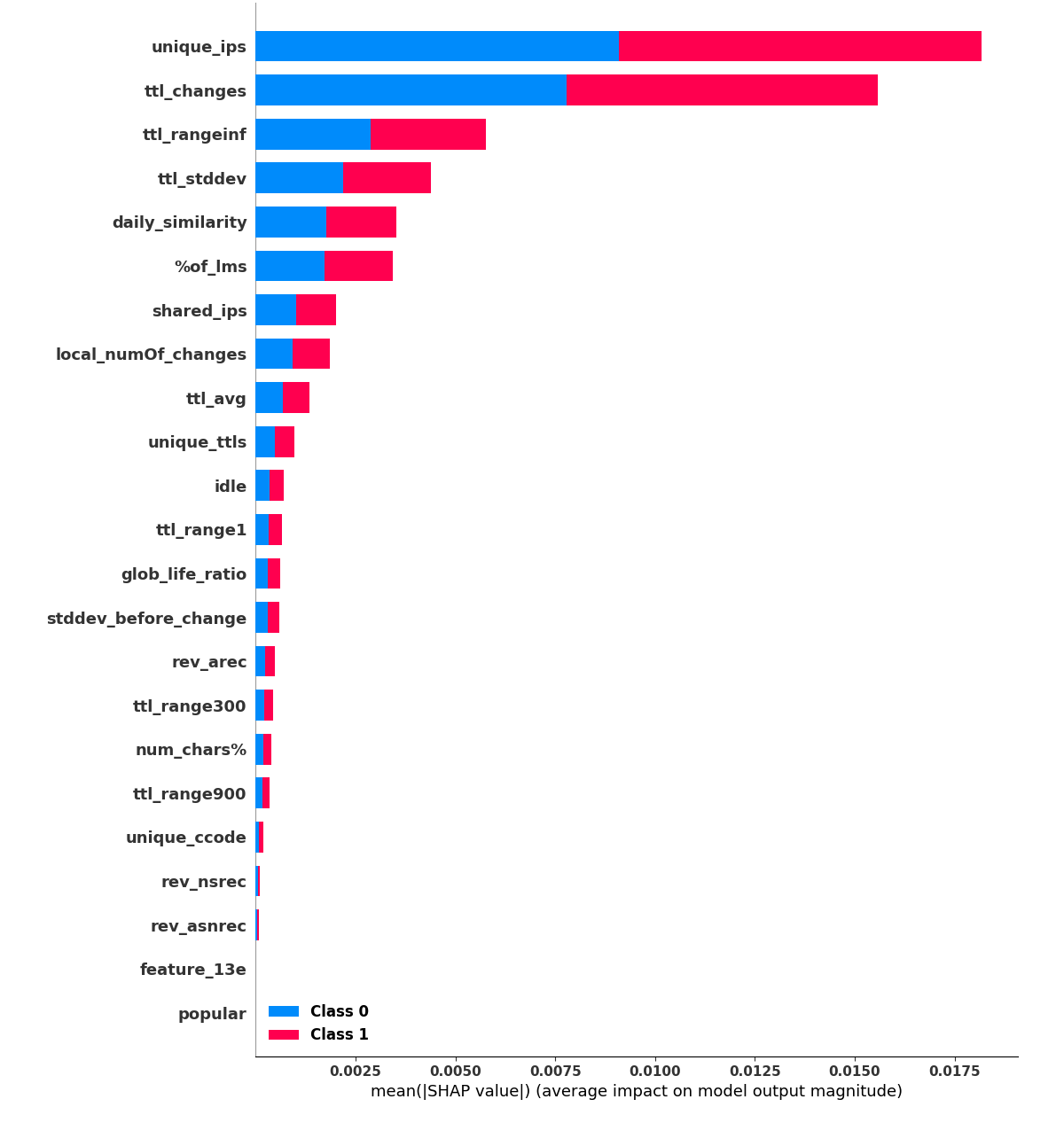}}
\qquad
\subfloat[SHAP summary plot of feature contributions on RANDOM FOREST classifier.]{\label{fig:shap_rf}\includegraphics[width=0.475\linewidth]{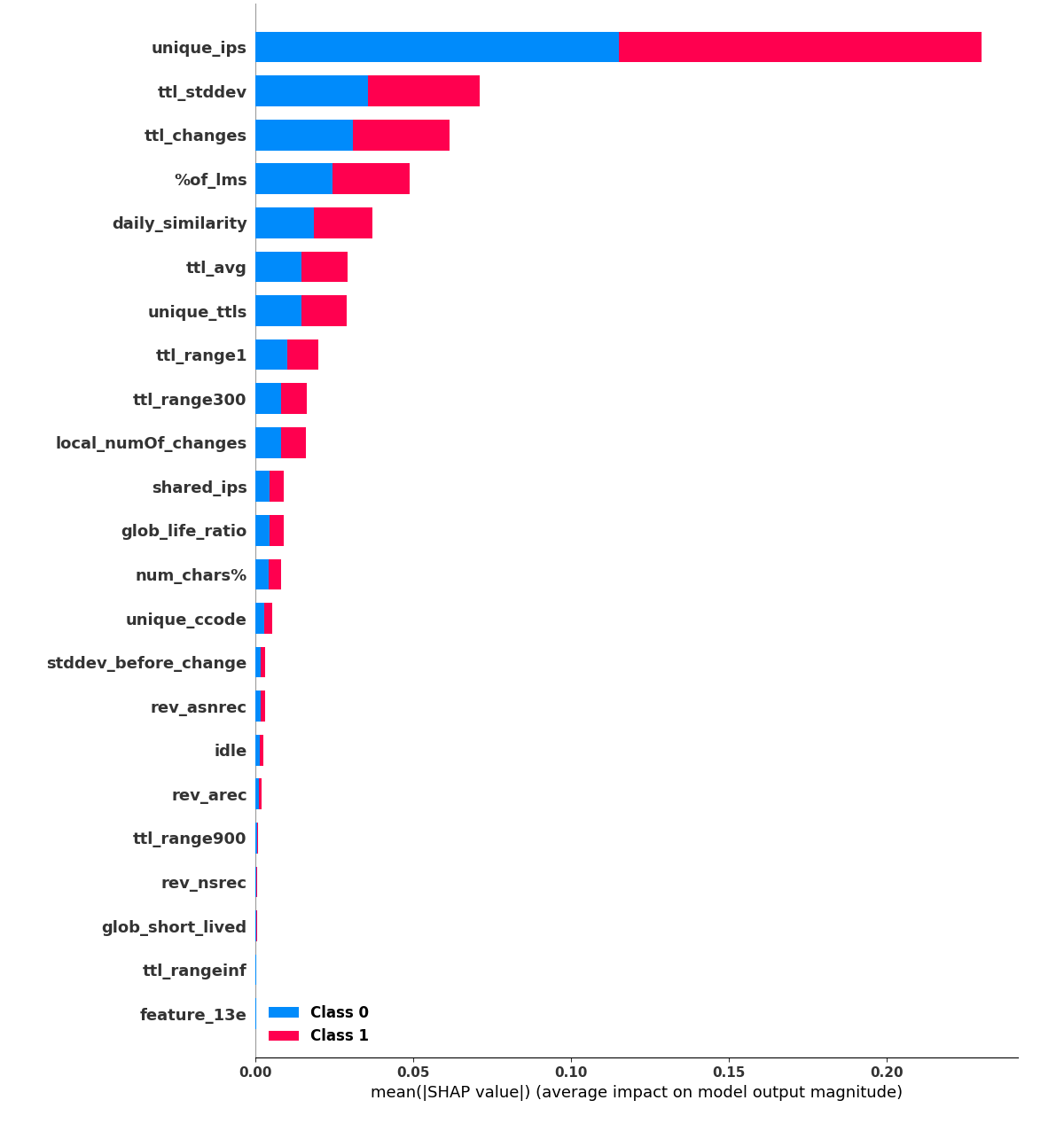}} 
\caption{Global summary plots for ADA (\textit{left}) and RF (\textit{right}) classifiers.}
\end{figure}

\myparagraph{Interpreting Global Explanations.} As pointed out in Section~\ref{sec:explainability}, different models can use features in different ways. 
Global explanations can uncover these behaviors and let the analyst be aware of the feature prioritization that a model brings. 
Some Decision-Tree based classifiers such as ADA and RF, respectively in the summary plots of Figure~\ref{fig:shap_ada} and Figure~\ref{fig:shap_rf}, share four out of the five top important features, which is likely to be a consequence of their ``similar" tree-based intrinsic nature. 
In both of them, \uniqueips notably brings the higher contribution. 
In Section~\ref{sect:appendix}, we show how other classifiers have a low magnitude provided by the \uniqueips feature, whilst prioritizing a diverse subset of TTL features.  
This furtherly shows how it is not possible to solely rely on statistical analysis to foresee the utilization of the features, as class separations that at first glance look either weak or strong can be subverted. 

\begin{figure}[htbp]  
\centering 
\subfloat[SHAP PDP on \uniqueips feature.]{\label{fig:pdp_uniqueips} \includegraphics[width=0.45\linewidth]{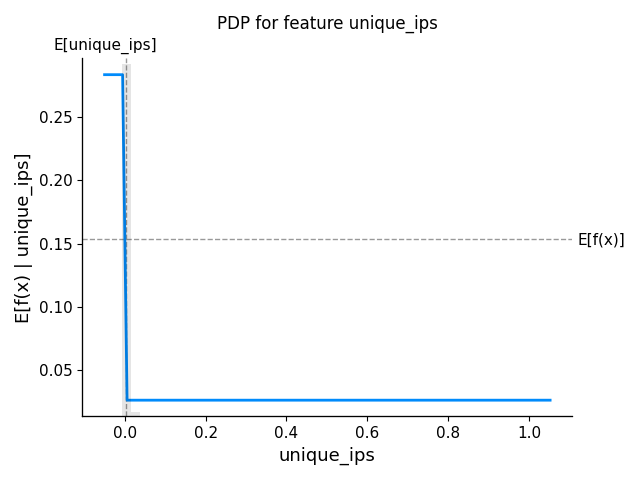}}
\qquad
\subfloat[SHAP PDP on \numchars feature.]{\label{fig:pdp_numchars}\includegraphics[width=0.45\linewidth]{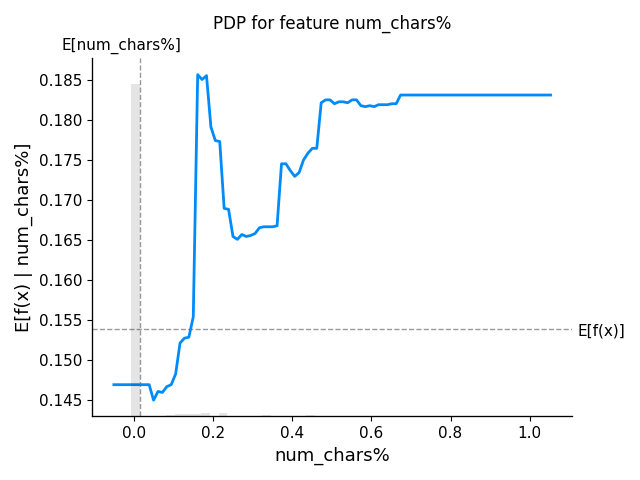}}

\caption{Partial Dependence plots for \uniqueips (\textit{left}) and \numchars (\textit{right})}
\end{figure}

Partial Dependence Plots show the marginal effect of a single feature globally on the predictions. Considering a trained classifier (RF in the case of this analysis) and a background distribution, through SHAP we can assess how the considered feature contributes to classifying the background samples over their values. An advisable security-related use of these plots can be to employ a background distribution of malicious samples, thus analyzing to which extent the feature values contribute to classifying the sample as malicious. The following plots (after normalization) have been made using a background distribution of 1000 malicious samples on the RF classifier. Figure~\ref{fig:pdp_uniqueips} shows the PDP of the strongest feature of the model. The plot tends to be a ``gentle" step, producing the highest contribution on very low feature values and the lowest with values going just subtly over the threshold. Very similar behavior to the one of the \uniqueips feature is shown in the number of changes in the TTL, depicted in Figure~\ref{fig:pdp_ttlchanges} in Section~\ref{sect:appendix}. These plots help, for example, understanding the extent to which features contribute to the classification of the domain as malicious and possibly setting policies and restrictions based on simply tweakable features, such as the \numchars in Figure~\ref{fig:pdp_numchars}. 
In this plot, we can understand how a high rate of numerical characters leads to a solid contribution to the prediction of the domain as malicious. Likewise, it is surprising that a 20\% rate of numerical characters in the domain string leads to an even bigger magnitude, which can be caused by the relevant presence of some malware families not having numbers in their ``regex". In this case, usage of the proposed security policies for a system hosting an EXPOSURE-like system would be to allow domain names with numerical characters comprised in between the 20\%-40\% range. 

\begin{figure}[htbp]  
\centering 
\subfloat[Force plot of the benign domain sample \texttt{spring.io}, correctly classified as benign.]{\label{fig:spring}\includegraphics[width=\linewidth]{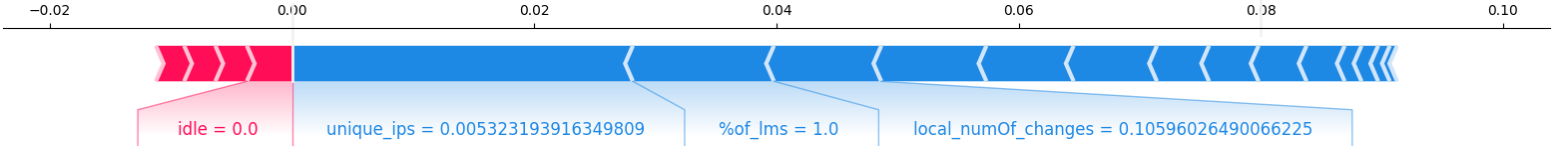}} \\
\subfloat[Force plot of the malicious domain sample \texttt{fgc.se}, misclassified as benign.]{\label{fig:fgc}\includegraphics[width=\linewidth]{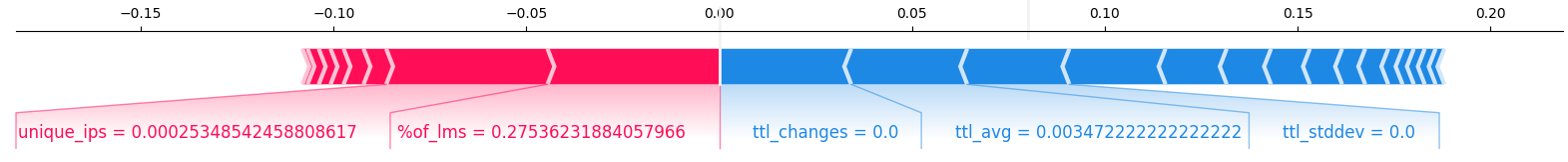}} \\
\subfloat[Force plot of the benign domain sample \texttt{topeleven.com}, misclassified as malicious.]{\label{fig:topeleven}\includegraphics[width=\linewidth]{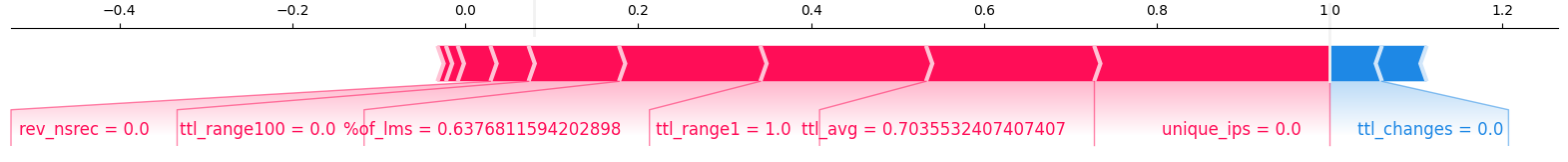}} 
\caption{Local explanations on three domains from the dataset.}\label{fig:force_plots}
\end{figure}

\myparagraph{Magnifying the Behavior with Local Explanations.}
Force plots and local explanations can be considered the local version of summary plots, showing how features contribute locally to the sample. As Figure~\ref{fig:spring} shows, the features correctly lead the RF model classification of the domain \texttt{spring.io} as benign. It is not surprising to see a 0 value of \uniqueips forcing the classification towards malicious, but the rest of the features gently move the prediction towards the \textit{benign} class. For malicious samples, instead, the latter plays a major role incorrectly classifying some malicious samples like \texttt{mobile.de} and \texttt{qcx.nl}, as additionally shown in the Appendix.
In general, the magnitude of the features sticks to what is shown by the summary plot of Figure~\ref{fig:shap_rf} for samples correctly classified. 
To understand instead which features are leading the model to misclassify a sample, Figure~\ref{fig:fgc} shows the force plot for the domain \texttt{fgc.es}, blacklisted but yet misclassified by our system as \textit{benign}. 
The \uniqueips and \lms features correctly move the prediction towards the \textit{malicious} class, but the values of the TTL-based features deceive the classifier. Figure~\ref{fig:topeleven}, on the other hand, shows how \uniqueips and TTL values deviate the prediction of the benign domain towards the malicious class. 

\myparagraph{Summary of the Results.} 
As a result of the presented experiments, we can reflect on the issues of feature management and hypothetical counteractions. 
We understand the features distribution, correlation, and how the DGAs in our traffic tend to behave from statistical analysis. 
However, the global explanations can turn the table and quantify how the model perceives the features. 
Finally, we can see how features drive the sample's prediction via local explanations. 
This ensemble of analysis makes us notice how the overall feature prioritization depends on both the model used and the considered data, which further proves how context-dependent such systems' behavior can be. 
Hence, an Explainability analysis should always be used to better portray the big picture of both systems and employed data. 
In our case, the big picture has led to a more prominent analysis of the EXPOSURE system. 
The TTL features, the subject of this analysis, sum up to 37.5\% of the entire feature set, being 9 out of 24. 
It turns out from our analysis of the explanations that they contribute massively to the misclassification of several samples (such as Figure~\ref{fig:fgc} and Figure~\ref{fig:topeleven}) as they cover high-magnitude roles in the summary plots of Figure~\ref{fig:shap_ada} and Figure~\ref{fig:shap_rf} (which are the best classifiers overall). 
This makes this feature set highly powerful for the whole system, yet in its power also lies a crucial problem. 
Namely, attackers can manipulate this feature, being completely free to tune the TTL and balance their caching time (\ie, the likelihood of being detected) with the chance of evading the classification of the system. 
Having such a relevant portion of the feature set reserved for values that can be somehow crafted directly by the attackers, it can serve as a significant stepping-stone for attackers. 
Furthermore, if deployed on a system devised to manage critical resources, besides evading, some part of the feature set can be overridden by the context. 
Some works like~\cite{vlajic_role_2012} point out how security-sensitive systems, \eg, banking applications, should indeed carefully set their DNS TTL to a low value. 
The scope of these assumptions is that a botnet/DGA detector cannot solely rely on accuracy metrics to establish its efficiency, in that analysts also need to be aware of the model, the data, and the context therein. 
Explanations can give a huge and crucial hand in this regard, helping prevent major issues from happening and allowing debugging of the model. 
Considering our system, through explanations, we have seen how dangerously influential TTL-based features are in most of the models. And considering their extensive use in the feature set, appropriate security measures should be taken (\eg, reducing their number like for Name-Based features, which are just easily adjustable as well but sum up to only the 8\%). 
We firmly believe that through explanations we can rapidly enhance the usage and trust in AI, as companies can look at such security systems from a human-readable perspective, and model biases can be analyzed and studied.

\section{Related Work}\label{sect:related_works}

\myparagraph{DNS Analysis.} Several promising works have striven to tackle botnet/DGA detection during the last decade, often showing innovative DNS passive features and methodologies. Some notable works, besides EXPOSURE, have been Notos~\cite{antonakakis2010building}, where Antonakakis et al. created the first relevant and efficient reputation system for domains from various data sources. Pleiades~\cite{antonakakis2012throw}, where the authors focused on NXDOMAIN records to both cluster domains and classify DGAs by looking at the strings association. Finally, in FANCI~\cite{Schueppen2018}, Schueppen et al. developed a detector based on a small feature set such as the EXPOSURE one, though focusing only on NXDOMAIN passive data. All of these works have reached comparable performances in different settings.
None of them, though, has focused their interest on the explainability of such a critical application. The only works to have addressed such problems focused on multiclass classification problems with deep learning approaches, thus classifying the malicious domains with family pairing. In~\cite{9347410}, Becker et al. proposed a visual analytics system for Deep Learning (DL) models, providing graphical insights on statistical properties of the domain name string. 
Drichel et al., in~\cite{inproceedingsdrichel1}, briefly highlighted some string-wise interpretations for DL models starting from the misclassified samples. In contrast, in~\cite{inproceedingsdrichel2}, Drichel et al. proposed feature-based classifiers based on string features for multiclass classification, with the purpose of improving explainability.
Firstly, none of these three works focused on passive DNS data, choosing string-based features to ease the computational burden. Secondly, none developed explanatory analysis, rather focusing on how the model could be made more explainable or at most on how to visualize a few string patterns from a DL model. 
Our work focuses on passive DNS traffic data, analyzing features from a comprehensive viewpoint and not limiting them to the human-readable string features. Additionally, we propose both local and global explanations, concretely enhancing the awareness of how a model behaves in such a context.

\myparagraph{Explainability Techniques.} In~\cite{ribeiro_why_2016}, Ribeiro et al. proposed LIME (Local Interpretable Model-Agnostic Explanations), an explainability method conceived as a local model learning and approximating around the prediction.
Despite its wide use, several concerns about stability and consistency have been addressed towards LIME~\cite{lundberg_unified_2017, zhou_s-lime_2021}. 
Considering that SHAP is a more reliable tool, we have driven our choice towards its use in our explainability work.

\section{Conclusions and Future Work} \label{sec:conclusions}
In this work, we proposed an explanatory analysis of ML classifiers devised for botnet/DGA detection. Starting from the implementation of the EXPOSURE feature set on our traffic data, we have shown how from prior statistical assumptions on the malware behavior within the network, a model can interpret features in its way globally, thus prioritizing certain features rather than others that were prevented, also demonstrating how different models can have a different feature conception, to which eventually we analysts should adapt and debug accordingly. Locally, we have seen how certain features can contribute and how explanations can make the analysts and users aware of the single decisions and motivations behind misclassified samples. Through these analyses, we raised concerns about how the feature and model can be biased by the context in which the systems are both trained and deployed. And our analysis makes the comprehension of such contexts move fast forward towards favoring the employment of such systems, as they can be firstly interpreted and adapted and subsequently accepted. In this regard, several advances of this work can be developed aiming at fairness and legal regularization of the detectors through explanations and, if possible, bringing them into debugging/pipelining processes to obtain an efficient and explainable system. Additionally, they can be instrumental when humans want to be involved in the decision-making. All in all, this work demonstrated how powerful explanations can be and how security, debugging, interpretability, and fairness can be brought to the next level by the application of ML to detection, where security has to be assessed and interpreted through the process chain.

\section*{Acknowledgments}

This work has been partly supported by the PRIN 2017 project RexLearn, funded by the Italian Ministry of Education, University and Research (grant no. 2017TWNMH2); and by the project TESTABLE (grant no. 101019206), under the EU’s H2020 research and innovation programme.

\bibliographystyle{unsrt}
\bibliography{bibliography}

\newpage
\section{Appendix}\label{sect:appendix}
In this additional section, we show several plots that have been cited in the previous sections and that we believe can support the comprehension of the work.

\myparagraph{Grid Search Results.} Using the Scikit-Learn Python suite, we optimized the parameters through the GridSearchCV API.
The results of the optimization have been reported for completeness in Listing~\ref{lst:ls_grid}.

\begin{figure}[htbp] 
\centering
\includegraphics[width=0.5\linewidth]{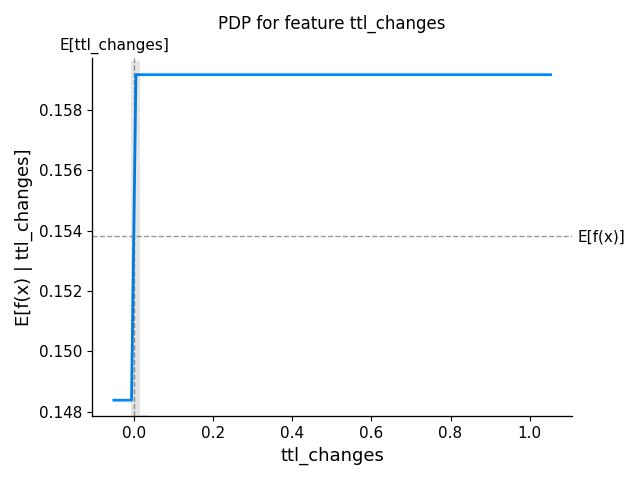} 
\caption{SHAP PDP on $ttl\_changes$ feature.}
\label{fig:pdp_ttlchanges} 
\end{figure}

\begin{figure}[htbp] 
\centering
\includegraphics[width=\linewidth]{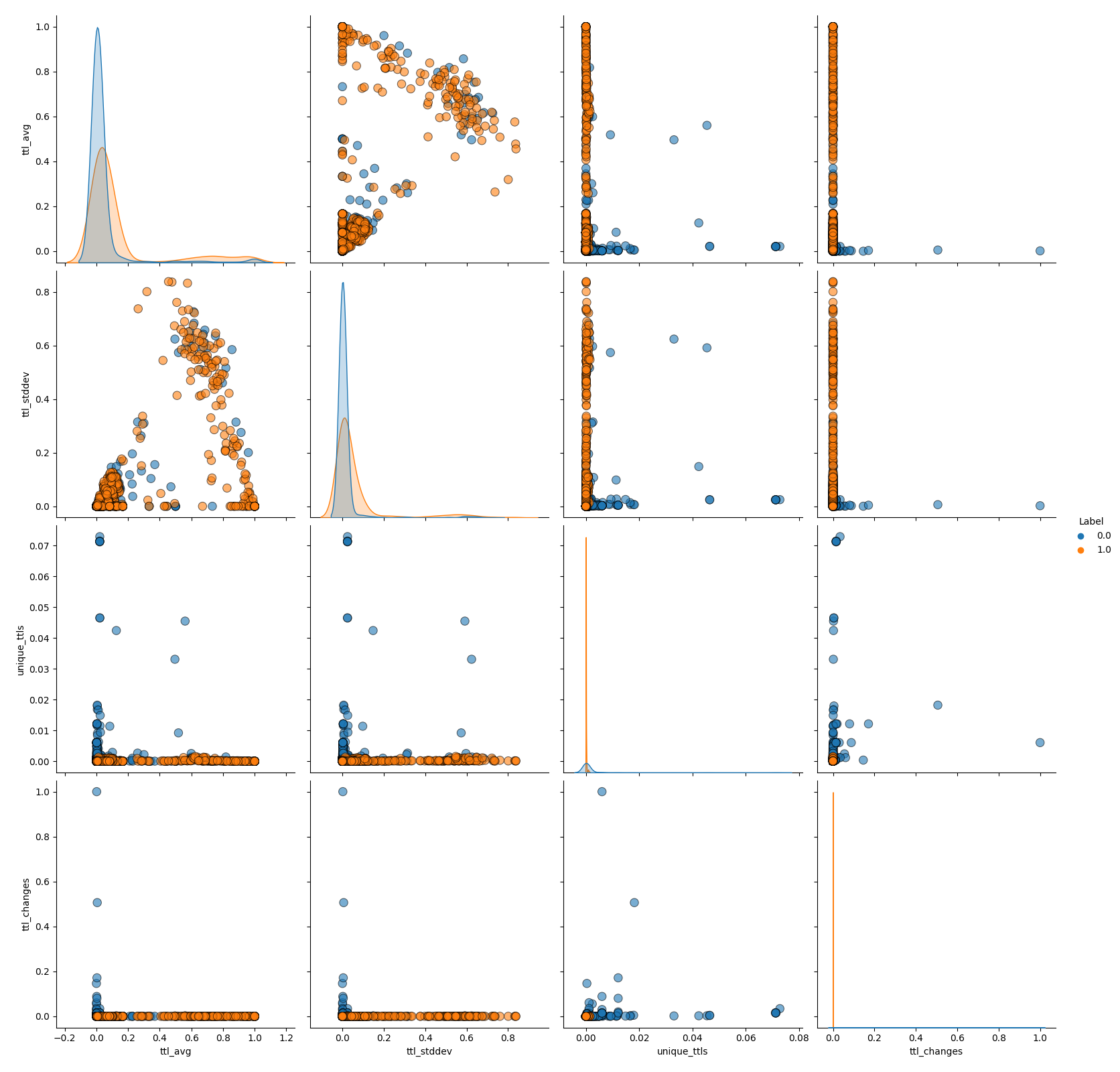} 
\caption{Correlation between the first four TTL-based features.}
\label{fig:seabornttl} 
\end{figure}

\myparagraph{TTL Features plots.} 
Having focused the discussion of the explainability analysis almost entirely on the TTL features, there are some additional plots that can point out interesting behaviors, such as Figure~\ref{fig:seabornttl}, which shows the statistical analysis of the first 4 TTL-based features. In Figure~\ref{fig:pdp_ttlchanges} instead, we can see how low changes in the TTL values mean a low contribution to the classification of the sample as malicious and vice versa. 

\begin{figure}[htbp] 
\centering
\includegraphics[width=0.7\linewidth]{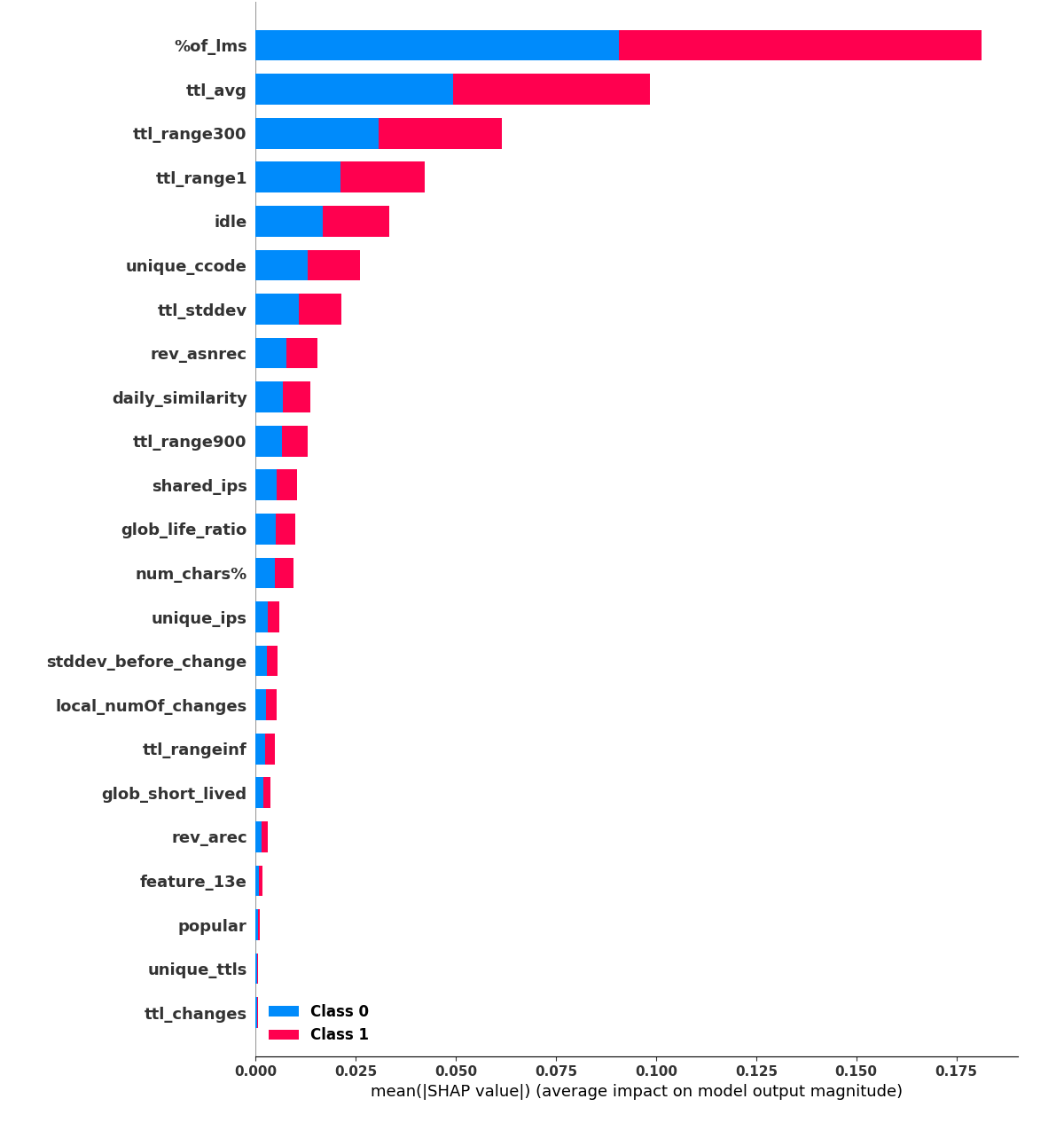} 
\caption{SHAP summary plot of feature contributions on KNN classifier.}
\label{fig:shap_knn} 
\end{figure}

\begin{figure}[htbp] 
\centering
\includegraphics[width=0.7\linewidth]{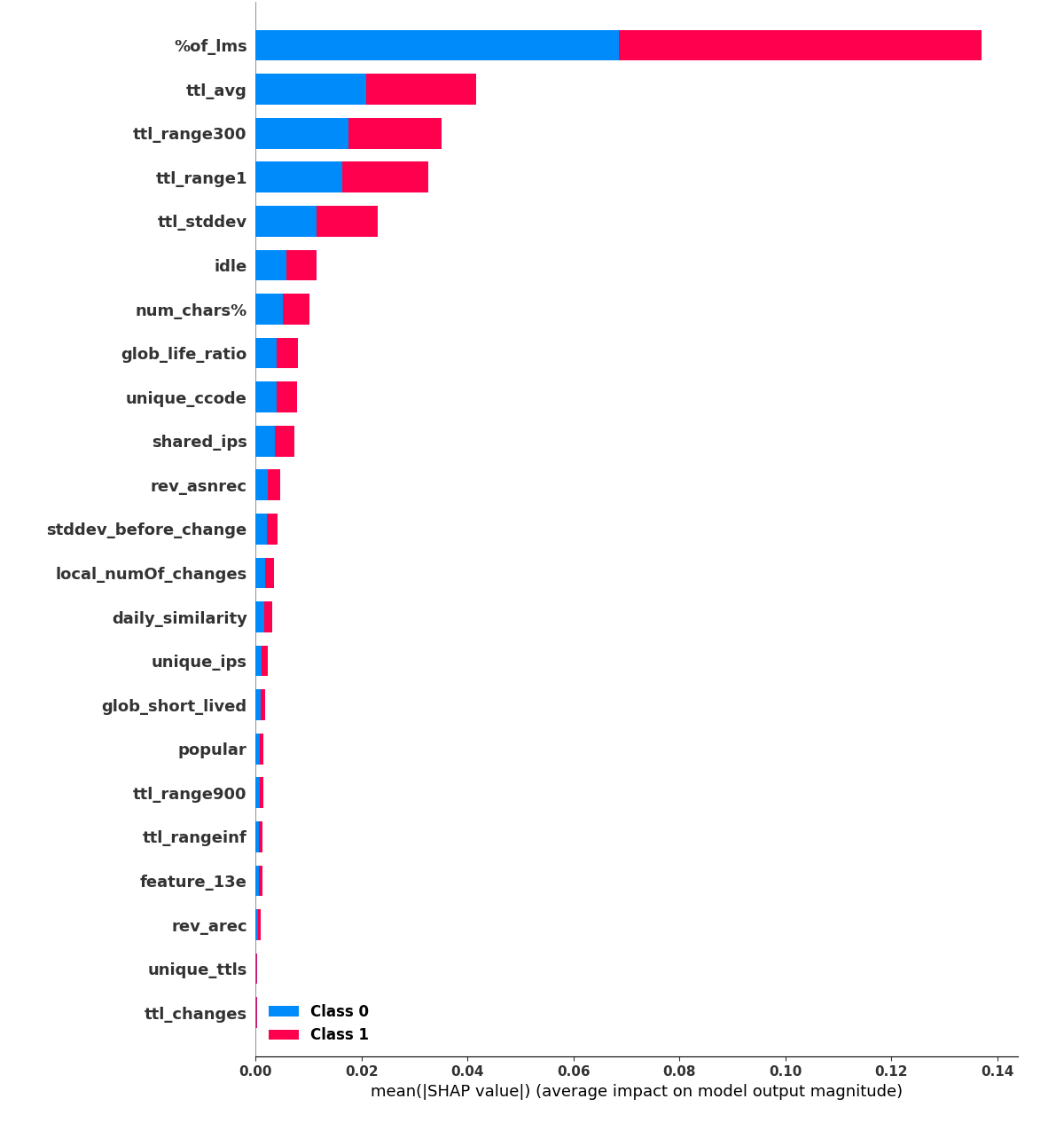} 
\caption{SHAP summary plot of feature contributions on SVC classifier.}
\label{fig:shap_svc} 
\end{figure}

\myparagraph{Additional Summary Plots.}
Figure~\ref{fig:shap_knn} shows how \uniqueips are much less considered than the TTL-based features by the KNN classifier, which again shows how models are as diverse as they are. The same goes for the SVC classifier in Figure~\ref{fig:shap_svc}, which once again does not employ the \uniqueips feature as much as the Decision-Tree based classifiers do.

\begin{figure}[htbp]  
\centering 
\subfloat[Force plot of the benign domain sample \texttt{mobile.de}, correctly classified as benign.]{\label{fig:mobilede}\includegraphics[width=\linewidth]{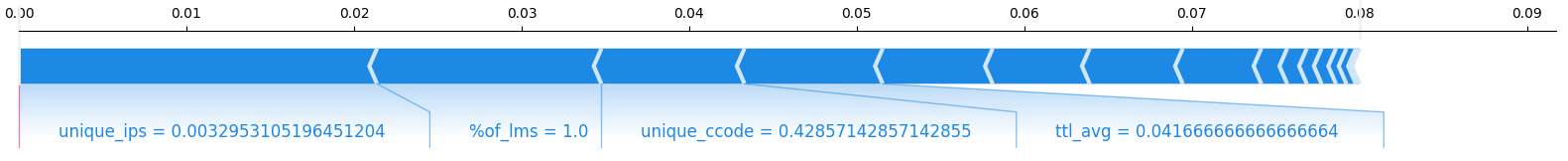}} \\
\subfloat[Force plot of the malicious domain sample \texttt{qcx.nl}, correctly classified as malicious.]{\label{fig:qcx}\includegraphics[width=\linewidth]{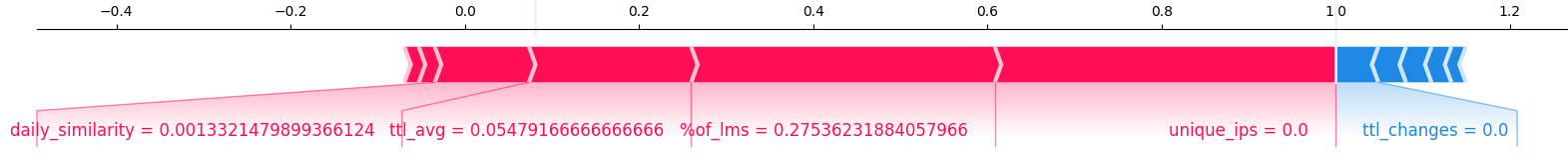}} \\
\subfloat[Force plot of the malicious domain sample \texttt{szx.pw}, misclassified as benign.]{\label{fig:szx}\includegraphics[width=\linewidth]{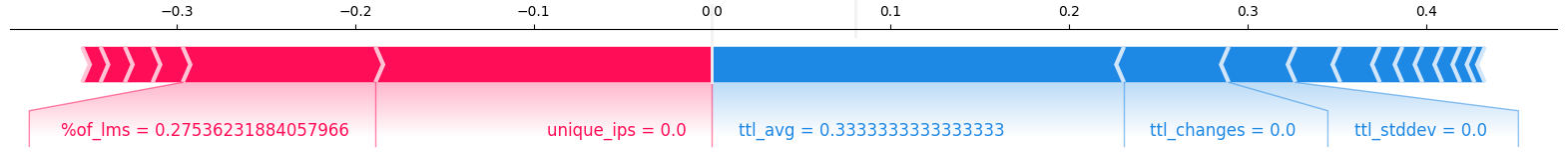}} 
\caption{Local explanations on other three domains from the dataset.}\label{fig:force_plots_more}
\end{figure}

\myparagraph{Additional Force Plots.} The plots of Figure~\ref{fig:force_plots_more} show a variety of samples either correctly classified or misclassified by the RF model, demonstrating practically how the most relevant features can play a major role in any classification scenario, either in the wrong or correct way.

\begin{lstlisting}[caption={Grid Search Results},label={lst:ls_grid},language=Python]
# RANDOM FOREST parameters
{'criterion': 'entropy', 'max_depth': 20, 'n_estimators': 125}
# ADA-BOOST parameters 
{'algorithm': 'SAMME', 'n_estimators': 175}
# K-NEAREST NEIGHBORS parameters
{'n_neighbors': 13, 'weights': 'distance'}
# DECISION TREE parameters
{'criterion': 'gini', 'max_depth': 5}
# SVC RBF parameters 
{'C': 297.63514416313194, 'gamma': 0.6951927961775591}
\end{lstlisting}

\end{document}